\documentclass[useAMS,usenatbib,usegraphicx]{mn2e}
\title[Type Ia Supernovae vs. PSCz]{A Comparison of Local SNIa with
the IRAS PSCz Gravity Field}
\author[Radburn-Smith et al]{D.~J.~Radburn-Smith$^1$, J.~R.~Lucey$^1$
  and M.~J.~Hudson$^2$\\
$^1$Department of Physics, University of Durham, South Road, Durham
  DH1 3LE, UK\\
$^2$Department of Physics, University of Waterloo, ON, N2L 3G1, Canada}

\begin{document}

\date{Accepted 2004 September 22. Received 2004 August 20; in original form 2004 June 18}

\pagerange{\pageref{firstpage}--\pageref{lastpage}}
\pubyear{2004}

\maketitle

\label{firstpage}

\begin{abstract}
We compare the measured peculiar velocities of 98 local
($<150~h^{-1}$~Mpc) type Ia supernovae with predictions derived from
the PSCz survey\nocite{bra99}. There is excellent agreement between
the two datasets with a best fit $\beta_I$ ($=\Omega_m^{0.6}/b_I$) of
$0.55\pm0.06$. Subsets of the supernovae dataset are further analysed
and the above result is found to be robust with respect to culls by
distance, host-galaxy extinction and to the choice of reference frame
in which the analysis is carried out. Alternative methods of
determining $\beta_I$ including density-density comparisons, dipole
measurements and WMAP-based results are also discussed. We conclude
that most recent determinations are consistent with a value of
$\beta_I=0.5$.
\end{abstract}

\begin{keywords}
cosmology: observations -- cosmological parameters -- dark matter --
galaxies: distances and redshifts -- large-scale structure of Universe
\end{keywords}

\section{Introduction}

Peculiar motion studies are a powerful tool for examining the
underlying mass distribution of the local universe. In the linear
regime, where density fluctuations are small, the mass over-density,
$\delta_m$, can be related to the fluctuation in galaxy
number-density, $\delta_g$, via $\delta_g(r)=b_{}\delta_m(r)$ where
$b$ is the linear bias parameter. This bias parameter together with
the cosmological mass density parameter, $\Omega_m$, can be used to
predict peculiar velocity fields from all-sky redshift surveys via the
equation
\begin{equation}
{\bmath{v(r)}}=\frac{\beta}{4\pi}\int\delta_g({\bmath{r'}})\frac{{\bmath{r'-r}}}{\mid{\bmath{r'-r}}\mid^3}d^3{\bmath{r'}}
\label{eq:1}
\end{equation}
The dimensionless quantity $\beta$ ($=\Omega_m^{0.6}/b$) scales
linearly with the predicted velocities and is the only free parameter
of the model \citep{peb80}. Hence $\beta$ can be directly determined from
the comparison of measured peculiar motions with predictions made from
galaxy density fields.

All-sky galaxy samples derived from {\em{IRAS}} satellite data have
been used extensively to map the local density field. Currently the
most complete redshift survey of {\em{IRAS}} sources is provided by
the PSCz \citep{sau00}. This survey consists of redshifts for 15,411
galaxies uniformly distributed over 84.1\% of the sky with a median
redshift of 8500 km~s$^{-1}$. The PSCz survey's depth, excellent sky
coverage and density allow for the reliable mapping of the
distribution of galaxies in the local universe. Several independent
determinations of the PSCz density and velocity fields have therefore
been made; most notably by \cite{bra99}, \cite{sch99} and
\cite{row00}. Recent comparisons of these fields with peculiar velocity
measurements typically yield values of $\beta_{I}$ in the range
0.4 - 0.6 \citep[see][]{zar02}.

A significant source of error in determining $\beta$ arises from the
uncertainty in the peculiar velocity measurements. Galaxy distance
estimates from the Tully-Fisher and Fundamental Plane relations are
subject to errors that are typically $\sim20\%$ per galaxy. At depths
greater than $\sim50~h^{-1}$~Mpc this is considerably larger than the
peculiar velocities of the individual galaxies. With distance errors
less than $10\%$, Type Ia supernovae (SNIa) are less susceptible to
inhomogeneous Malmquist bias \citep{hud94} and hence offer an important
alternative probe of the local velocity field. An early attempt to use
SNIa was carried out by \cite{rie97} who compared the peculiar
velocities of 24 SNIa with the velocity fields predicted from the
1.2~Jy IRAS redshift survey \citep{fis95} and the Optical Redshift
Survey \citep{san95,bak98}. They derived $\beta_I=0.4\pm0.15$ and
$\beta_O=0.3\pm0.1$ respectively, with the relatively large error
resulting from the small sample size.

\cite{ton03} have recently produced a homogenized compendium of 230
SNIa for constraining cosmological quantities. The release of this
compendium presents a new opportunity to measure $\beta$ with a
significantly smaller error. In this paper we compare peculiar
velocities measured by the local SNIa in the \cite{ton03} sample to
the peculiar velocity field derived from the smoothed PSCz density
field as determined by \cite{bra99}.

We briefly describe the \cite{ton03} compilation of SNIa in
Section~\ref{sn} and review the derivation of the PSCz velocity field
in Section~\ref{pscz}. In Section~\ref{beta} we describe the
derivation of $\beta_I$. The robustness of this result is analysed in
Section~\ref{robust} and finally, in Section~\ref{conc} we discuss the
results and present our conclusions. 

\section{The SNIa dataset}\label{sn}

The \cite{ton03} dataset is a homogenized compendium of 230 SNIa
compiled from many recent studies. Most notably from the \cite{jha02},
\cite{per99}, \cite{ham96}, \cite{rie99} and \cite{ger04} datasets, which
comprise the majority of the data. Using a variety of fitting
techniques such as MLCS (Riess et al. 1998 and the work of Jha and
  collaborators)\nocite{rie98} and dm15 \citep{ger01}, \cite{ton03} have
re-calculated the relative SNIa distances where the original
photometric data is available. The systematic offsets of each dataset
were reduced by minimising the differences between all pairs of
datasets where overlaps exist. The residuals of this fitting procedure
are 0.02 mag or better for the majority of the samples. Table~15 of
\cite{ton03} lists the redshift ($\log{cz}$), luminosity distance
($\log{dH_0}$), distance error and host galaxy $V$-band extinction
($A_V$) for each SNIa.

\cite{ton03} fix the zero point of nearby SNIa ($0.01<z<0.1$) by assuming
an `empty universe' ($\Omega_{m} =0, \Omega_{\Lambda} = 0$)
cosmology.  For our analysis, we have converted the \cite{ton03}
quoted distances to a $\Lambda$CDM cosmology ($\Omega_{\Lambda}=0.7,
\Omega_m=0.3$).  However, the derived $\beta_I$ is unaffected by the
choice of cosmology.

In this paper we only consider the 107 SNIa that lie within 150 $h^{-1}$
Mpc as the PSCz density field is incomplete at greater distances for
all galactic latitudes \citep{bra99}.  We further restrict the sample
to SNIa with extinctions $A_V < 1.0$ mags, for reasons discussed
below. These selection criteria leave 98 SNIa, which we refer to as the
``default sample''.  The median distance error for this local SNIa
sample is $\sim8\%$.

\section{The PSC$z$ Velocity Field Model}\label{pscz}

\cite{bra99} used the PSCz redshift survey to determine the density
and peculiar velocity fields in real space in a self-consistent way by
using equation (\ref{eq:1}) under the assumption that mass follows the
number density of {\em{IRAS}} galaxies. These fields are smoothed with
a Gaussian filter of radius $5~h^{-1}$~Mpc. Analysis by \cite*{ber00}
indicates that this smoothing radius should yield unbiased results for
$\beta_I$. In an independent analysis, \cite{sch99} derived the PSCz
velocity and density fields by using a Fourier-Bessel approach. They
found the resulting fields to be consistent with the \cite{bra99}
fields used here.

The integral in equation (\ref{eq:1}) extends over all space. The PSCz
survey, however, does not extend to infinite depth, nor does it have
data in the Zone of Avoidance (ZoA). For the ZoA, \cite{bra99} have
implemented a similar approach to that of \cite{yah91} by dividing the
region ($|b|\le8\degr$) into bins of $10\degr$ latitude by
1000~km~s$^{-1}$. These bins are then populated with enough synthetic
galaxies to reflect the number density of the corresponding bins at
greater $|b|$. The systematic effect on the derived value of $\beta_I$
due to this interpolation procedure can be estimated from the results
of \cite{hud94b}. He compared $\beta$ values derived from an
optically-selected density field with a larger ZoA ($|b|\le12\degr$)
using different techniques to account for the missing structure. Only
an 8\% difference was observed between the $\beta$ value derived from
the interpolated density field and that derived from a density field
in which the ZOA was assumed to be at average density. Since the
average density assumption is rather extreme, this result may be taken
as an upper limit on the systematic uncertainty. Therefore, as the
PSCz ZoA is only two-thirds the thickness of this ZoA, we might expect
a systematic uncertainty on our result of the order 5\%. This is
considerably smaller than our random errors.

As stated previously we have truncated the PSCz velocity field at
150~$h^{-1}$~Mpc due to increasing shot noise. Sources beyond this
depth, however, may still contribute to the LG's motion. Because the
statistical weight of the SNIa sample is dominated by nearby objects
these external contributions can be modelled as a dipole term. For
peculiar velocity comparisons in the LG frame this dipole term cancels
out as the motions of the LG and SNIa are affected in the same
way. LG-frame comparisons assume, however, that the LG's motion is
exactly given by linear theory. In practice, the LG is expected to
exhibit a nonlinear `thermal' component to its velocity that is not
well modelled by linear theory. An alternative to the LG-frame
comparison is to omit the LG from the analysis entirely. This can be
achieved by fitting the SNIa peculiar velocities in the CMB frame with
an additional dipole component to allow for contributions not included
in the PSCz density field. Ideally, analyses in both these frames
should produce similar results. However, due to the larger uncertainty
in the CMB analysis, we regard the LG result as a more reliable
solution.

\section{Determining $\beta_I$}\label{beta}

\begin{figure}
  \begin{center}
    \resizebox{90mm}{!}{\includegraphics{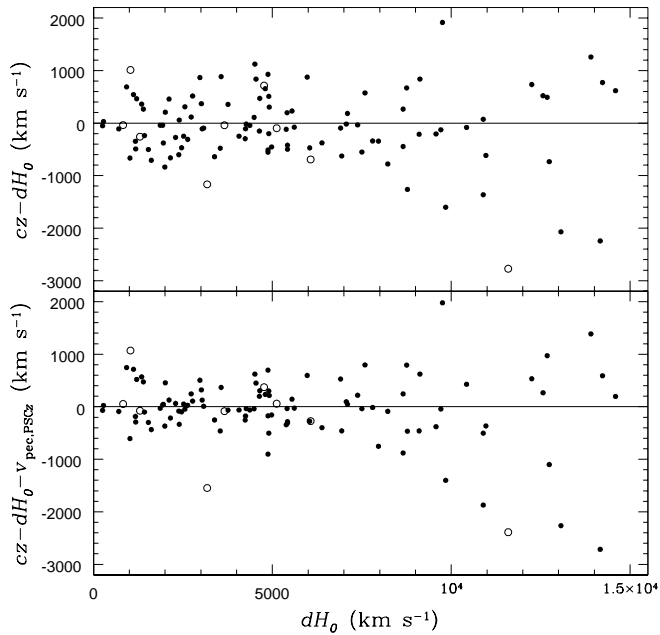}}
    \caption{The Hubble flow residuals for all 107 SNIa lying within 150
      $h^{-1}$ Mpc in the LG frame. The upper panel shows the original
      uncorrected data whilst the lower shows the data with the
      predicted PSCz peculiar velocities removed. Note the reduction
      in scatter, particularly in the distance range
      20-80~$h^{-1}$~Mpc. SNIa with host-galaxy extinctions
      $A_V\ge1.0$ are plotted as open circles whilst filled circles
      show the default sample used in this paper.
      }
    \label{hubble}
  \end{center}
\end{figure}

There is a very good agreement between the peculiar velocities
measured by the SNIa and predicted from the PSCz. This is shown in
Fig.~\ref{hubble} where the scatter around the Hubble flow before and
after the PSCz velocities for $\beta_I=0.5$ are removed. In the range
$20-80~h^{-1}$ Mpc, where the majority of SNIa lie, the removal of the
predicted PSCz peculiar velocities reduces the rms scatter around the
Hubble flow from 490~km~s$^{-1}$ to 390~km~s~$^{-1}$. In
Fig.~\ref{hubble}, nine SNIa with $A_V>1.0$ are plotted as open
circles, three of which are distinct outliers.  In our analysis we
have chosen to exclude these objects because we expect that their
errors are underestimated.

To determine $\beta_I$ in the LG frame we minimise the $\chi^2$
relation:
\begin{equation}
  \chi^2=\sum_{i} \Biggl(\frac{(v_{i,PSCz}-v_{i,SN})^2}{\sigma_{i,\rm
  cz}^2+\sigma_{i,d}^2}\Biggr)
\label{eq:beta}
\end{equation}
where $v_i$ is the peculiar velocity of the $i^{th}$ supernova,
$v_{i,PSCz}$ is the PSCz-predicted peculiar velocity which depends on
$\beta_I$ from (\ref{eq:1}), $\sigma_d$ is the distance error and
$\sigma_{\rm cz}$ incorporates both an estimate of the error in
redshift determination as well as errors in the PSCz predictions due
to shot noise or non-linear peculiar velocity contributions. 

Various studies have adopted different schemes for
$\sigma_{\rm{cz}}$. \cite{rie97}, adopt a value of 200~km~s$^{-1}$ for
all the SNIa, whilst \cite{bla99} use values of 150~km~s$^{-1}$ and
200~km~s$^{-1}$. However \cite{bla99} also account for the extra
velocity dispersion of cluster galaxies using two different
approaches. Their `Trial 1' method adds in quadrature an extra factor
of $\sigma_{\rm cl}(r)=\sigma_0/\sqrt{1+(r/r_0)^2}$ to $\sigma_{\rm
cz}$ where $\sigma_0=700\,(400)$~km~$^{-1}$ and $r_0=2\,(1)$ Mpc for
galaxies in Virgo (Fornax). Their `Trial 2' scheme uses the standard
$\sigma_{\rm cz}$ but resets the individual galaxy velocities for
group members to the group-average velocities as listed in
\cite{ton97} for 37 separate clusters. In our analysis we extended
both these techniques to account for galaxies which lie near one of
the X-ray selected clusters of the NOAO fundamental plane survey
\citep{smi04}.

Table~\ref{tab:cz} lists the derived $\beta_I$ values for these
different weightings for our default sample. The $1\sigma$ quoted
errors are calculated from bootstrap re-samples of the dataset. If the
nine $A_V>1.0$ SNIa had not been removed, the resulting $\chi^2$ would
be larger by $\sim40$. 

\begin{table}
 \caption{``Redshift error'', $\sigma_{\rm cz}$, comparison for the
  default sample of 98 SNIa in the LG frame. The errors have been
  determined from the 1$\sigma$ deviation in the distribution of the
  medians of 1000 bootstrap re-samples.}
 \label{tab:cz}
 \begin{center}
   \begin{tabular}{@{}lcc}
     \hline
     $\sigma_{\rm{cz}}^2$ (km$^2$~s$^{-2}$) & $\beta_I$ & $\chi^2$ \\
     \hline
     $150^2$ & $0.55\pm0.06$ & 167 \\
     $200^2$ & $0.54\pm0.06$ & 131 \\
     ${\bf 150^2+}{\bf \sigma}_{{\rm \bf{cl}}}^{\bf 2}$ {\textbf{`Trial 1'}} & ${\bf{0.55\pm0.06}}$ & {\bf{98}} \\
     $200^2+\sigma_{\rm cl}^2$ `Trial 1' & $0.54\pm0.06$ & 89\\
     $150^2$ `Trial 2' & $0.57\pm0.05$ & 97 \\
     $200^2$ `Trial 2' & $0.57\pm0.06$ & 88 \\  
    \hline
   \end{tabular}
 \end{center}
\end{table}

Increasing the redshift error $\sigma_{cz}$ for SNIa lying close to
nearby clusters has a sizeable effect on the $\chi^2$ but appears to
have no significant effect on the value of $\beta_I$. Overall, little
variation from the preferred value of $\beta_I=0.55\pm0.06$ is
observed and $\beta_I$ is effectively independent of the weighting
schemes used.

In order to determine $\beta_I$ in the CMB frame an extra dipole
component is added as an extra free parameter in the minimization of
equation~(\ref{eq:beta}). Using the default sample with
$\sigma_{\rm{cz}}$ given by `Trial 1' as
$\sqrt{150^2+\sigma_{\rm{cl}}^2}$, the best fit has
$\beta_I=0.48\pm0.09$ and $V_{dipole}=206\pm97$~km~s$^{-1}$ towards
$l=290\degr\pm25\degr$, $b=0\degr\pm18\degr$. This extra dipole
component is consistent with zero but is also consistent with the
value of $V_{dipole}=372\pm127$~km~s$^{-1}$ towards
$l=273\degr\pm17\degr$, $b=6\degr\pm15\degr$ as found by \cite{hud04}
for the SMAC sample. The calculated value of $\beta_I$ agrees well
with the result derived in the LG frame.

\begin{figure}
  \begin{center}
    \resizebox{90mm}{!}{\includegraphics{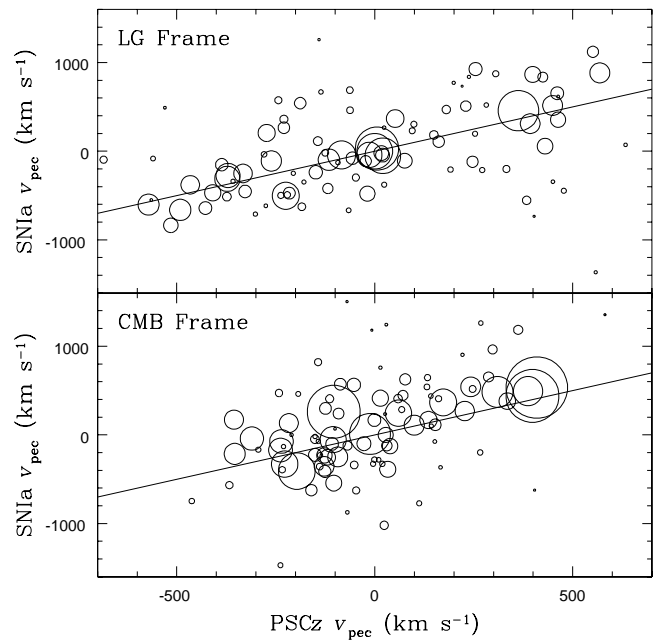}}

    \caption{Comparison of SNIa peculiar velocities to PSCz predicted
    peculiar velocities in the range $0~h^{-1}$~Mpc to
    $150~h^{-1}$~Mpc with $A_V<1.0$, $\sigma_{\rm
    cz}^2=150^2+\sigma_{\rm cl}^2$ and $\beta=0.55$. 
    The top panel shows comparisons in the LG frame, and the bottom
    panel shows the comparison in the CMB frame (without the extra
    dipole component). The size of the data point is inversely
    proportional to the total error
    ($\sigma=\sqrt{\sigma_{d}^2+\sigma_{cz}^2}$) on each SNIa. The
    smallest and largest circles correspond to values of
    $\sigma=1290$~km~s$^{-1}$ and 170~km~s$^{-1}$ respectively. The
    lines indicate a 1:1 ratio.}

    \label{fit}
  \end{center}
\end{figure}

The good agreement between the observed and predicted peculiar
velocities in both the LG and CMB frames is shown in
Fig.~\ref{fit}. If the peculiar velocities
predicted by the PSCz and observed from the SNIa are in exact
agreement for the chosen value of $\beta_I$, the SNIa would be
expected to lie along the 1:1 line. This trend is indeed observed. The
differences between the measured and predicted velocities are as
expected given the errors in both distance and velocity measurements,
i.e. the data is consistent with a reduced $\chi^2_{\nu}$ of
$\sim1$. Thus the two datasets agree exceptionally well.

\section{Robustness}\label{robust}

To assess the robustness of the derived $\beta_I$ we have examined
various sub-samples of the local SNIa dataset. Unless otherwise stated
all sub-samples use our default sample in the LG frame with
$\sigma_{\rm{cz}}=\sqrt{150^2+\sigma_{\rm{cl}}^2}$ determined using
the `Trial 1' approach. Table~\ref{tab:sample} lists the best fit
$\beta_I$ together with the associated $\chi^2$ for each sub-sample.

\begin{table*}
\begin{minipage}{130mm}
  \caption{$\beta_I$ dependency on sampling}
  \label{tab:sample}

  \begin{center}
    \begin{tabular}{lrcr} \hline \hline              
      Sample                   & No. SNIa & $\beta_I$ & Total
      $\chi^2_{min}$ \\ \hline
      0 $h^{-1}$ Mpc $<$ distance $<$ 150 $h^{-1}$ Mpc & 98 &
      $0.55\pm0.06$ & 98\\
      0 $h^{-1}$ Mpc $<$ distance $<$ 30 $h^{-1}$ Mpc & 31 &
      $0.55\pm0.07$ & 26\\
      30 $h^{-1}$ Mpc $<$ distance $<$ 150 $h^{-1}$ Mpc & 67 &
      $0.54\pm0.10$ & 74\\
      20 $h^{-1}$ Mpc $<$ distance $<$ 150 $h^{-1}$ Mpc & 80 &
      $0.55\pm0.07$ & 84\\
      40 $h^{-1}$ Mpc $<$ distance $<$ 150 $h^{-1}$ Mpc & 60 &
      $0.49\pm0.13$ & 67 \\
      0 $h^{-1}$ Mpc $<$ distance $<$ 100 $h^{-1}$ Mpc & 85 &
      $0.58\pm0.06$ & 78\\
      0 $h^{-1}$ Mpc $<$ distance $<$ 125 $h^{-1}$ Mpc & 90 &
      $0.56\pm0.06$ & 84\\
      No $A_V$ cull & 107 & $0.50\pm0.08$ & 141 \\
      $A_V<0.5$ & 80     & $0.57\pm0.06$ & 79   \\
      $A_V<0.3$ & 58     & $0.57\pm0.08$ & 57   \\
      CMB frame + dipole   & 98 & $0.48\pm0.09$ & 98 \\
      \hline
    \end{tabular}
  \end{center}
\end{minipage}
\end{table*}

Importantly, $\beta_I$ is found to be independent of the distance
range considered. Any derivation of $\beta$ is expected to be strongly
weighted by the very nearby SNIa where measurement errors are
smallest. Hence we have tested the dependency of our calculations on
SNIa at different distances by dividing the data into two distance
ranges. The position of this division is chosen such that the
bootstrap errors on each derived $\beta_I$ are of similar
magnitude. For a distance range of $0-30~h^{-1}$ Mpc we derive a value
of $\beta_I=0.55\pm0.07$ and for $30-150~h^{-1}$ Mpc,
$\beta_I=0.54\pm0.10$. Table~\ref{tab:sample} also includes a variety
of different distance ranges all of which yield similar values of
$\beta_I$ ($0.49<\beta<0.58$).

The determination of $\beta_I$ is also revealed to be independent of the
cull by host-galaxy extinction with $\beta_I$ varying by only
$\pm0.05$ for culls down to $A_V<0.3$. It is found that the reduced
$\chi_v^2$ is $\sim1$ for all culls of host-galaxy extinction
$<1.0$. Overall, for all the sub-samples considered, $\beta_I$ is found
to range by only 0.10.

Another source of bias which we do not account for in our analysis is
inhomogeneous Malmquist bias. Generally, not correcting for this will
lead to higher values of $\beta$. However, this bias scales with the
square of the distance error. Thus for the SNIa sample used here we
expect this bias to be considerably smaller than the random error in
$\beta_I$.

\begin{table*}
\begin{minipage}{110mm}
  \caption{Recent determinations of $\beta_I$ from velocity-velocity comparisons
}
  \label{tab:betas}
  \begin{center}
    \leavevmode
    \begin{tabular}{lcl} \hline \hline              
      Comparison & $\beta_I$ & Reference \\ \hline
      Mark III $vs.$ IRAS 1.2 Jy  & $0.50\pm0.10$   & \cite{dav96}  \\
      SNIa $vs.$ IRAS 1.2 Jy & $0.40\pm0.15$   & \cite{rie97}  \\
      SBF $vs.$ IRAS 1.2 Jy & $0.42^{+0.10}_{-0.06}$ & \cite{bla99} \\
      Mark III $vs.$ IRAS 1.2 Jy  & $0.50\pm0.04$   & \cite{wil98}  \\
      Mark III $vs.$ PSCz   & $0.60\pm0.10$   & \cite{sau99}  \\
      ENEAR $vs.$ PSCz      & $0.50\pm0.10$   & \cite{nus01}  \\
      SFI $vs.$ PSCz        & $0.42\pm0.04$   & \cite{bra01}  \\
      SEcat $vs.$ PSCz      & $0.51\pm0.06$   & \cite{zar02b}  \\
      \textbf{SNIa $vs.$ PSCz}   & \textbf{0.55} $\pm\textbf{ 0.06}$ &
      \textbf{This Study} \\
 \hline
    \end{tabular}
  \end{center}
\end{minipage}
\end{table*}

\section{Discussion}\label{conc}

Table~\ref{tab:betas} lists a representative set of recent
determinations of $\beta_I$ from comparisons of predicted and observed
peculiar velocities. Previously, the tightest constraints on
$\beta_I$ were from the merged spiral and elliptical peculiar velocity
samples such as Mark III \citep{wil97} and SECat \citep{zar00} as well as
the SBF sample of \cite{ton97}. This work adds a result from local
SNIa, a fourth independent data source of comparable statistical
power.  Recent comparisons of predicted and observed peculiar
velocities (`velocity-velocity'), including the result presented here,
all yield results consistent with a value of $\beta_I = 0.5$.

Some of the earliest estimates of $\beta$ were obtained by matching
the gravity at the LG to the measured CMB dipole.  While the LG has
the most accurate observed CMB-frame velocity, a weakness of this
method is that one needs to integrate the density field over all space
to obtain the predicted gravity at the LG. This contrasts
with the velocity-velocity comparison performed above in which
large-scale contributions to the predicted peculiar velocities either
drop out of the analysis (if the fits are performed in the LG frame)
or can be fitted independently of $\beta$ (if the fits are performed
in the CMB-frame). This degeneracy cannot be broken when using the LG
alone as one would be attempting to fit 4 parameters ($\beta$ and
three components of an external dipole) to 3 degrees of freedom (the
Cartesian components of the LG's CMB-frame motion).  Consequently, in
order to apply this method one needs either a deep, full-sky redshift
survey (so that the external dipole is known to be zero) or, failing
that, accurate estimates of the uncertainties arising from shot noise
at large distances and from incompleteness in the ZoA.  As an example
of the latter, \cite{hud04} have suggested, based on the ``Behind the
Plane'' extension of the PSCz \citep{sau00b}, that additional structure
in the ZoA beyond $100~h^{-1}$~Mpc may increase the PSCz dipole by
$\sim(170\pm85)$~km~s~$^{-1}$.  Until these issues are fully resolved,
$\beta$ determinations by this method remain subject to larger
systematic errors than velocity-velocity comparisons.

It is also possible to estimate $\beta$ by comparing the
density field inferred from redshift surveys, to the mass density
field constructed from peculiar velocity data. Such comparisons are
difficult because the mass density field is based on an inversion of
sparse and noisy peculiar velocity samples. The \mbox{POTENT}
reconstruction of the Mark III catalogue by \cite{sig98} yields
$\beta_I=0.89\pm0.12$. However, a new improved inversion method based
on an unbiased variant of the Weiner filter \citep{zar02b} finds
$\beta_I=0.57\pm0.12$, in good agreement with the velocity-velocity
results.

An alternative estimate of $\beta_I$ can be obtained from other
independent analyses not directly based on peculiar motion
studies. One noteworthy route is via the combination of parameters:
$\Omega_m^{0.6}\sigma_8$, where $\sigma_8$ is the rms amplitude of
mass fluctuations, $\delta_{m}$, averaged within a top-hat sphere of
8~$h^{-1}$~Mpc radius. This combination may be related to $\beta_I$ by
the dependence of $\sigma_{8,I}$, the number density fluctuation of
$IRAS$ galaxies, on the bias parameter $b_I$.
Since we are assuming linear biasing, $\delta_I = b_I \delta_m$ and
it follows that $\sigma_{8,I} = b_I \sigma_8$. We can thus write:
\begin{displaymath}
\beta_I=\frac{\Omega_m^{0.6}}{b_I}=\frac{\Omega_m^{0.6}\sigma_8}{\sigma_{8,I}}
\end{displaymath}
\cite{spe03} have used data from WMAP and other CMB and non-CMB
sources to derive a value of
$\Omega_m^{0.6}\sigma_8=0.38^{+0.04}_{-0.05}$.  By directly
integrating the PSCz power spectrum \cite{ham02} found
$\sigma_{8,I}=0.80\pm0.05$. Combining these two results gives
$\beta_I=0.48\pm0.06$. The good agreement of the results from all
these methods suggests that $\beta_I$ is now known at the 10\% level.

\section{Acknowledgements}

The authors would like to thank Enzo Branchini for providing the PSCz
velocity field. DJR-S thanks PPARC for a research studentship.  
MJH acknowledges support from NSERC and the ORDCF.

\bibliographystyle{mn2e}
\bibliography{snia_pscz}

\label{lastpage}

\end{document}